# Microscopic origin of the second law of thermodynamics

You-gang Feng

Department of Basic Science, College of  Science , Guizhou University
Guiyang, Cai Jia Guan, 550003, China

**Abstract**

We proved when random-variable fluctuations obey the central limit theorem the equality of the uncertainty relation corresponds to the thermodynamic equilibrium state. The inequality corresponds to the thermodynamic non-equilibrium state. The uncertainty relation is a quantum-mechanics expression of the second law of thermodynamics originated in wave-particle duality. Formulas of mean square-deviations changes adjusted by random fluctuations under the minimal uncertainty relation are obtained. Finally, an assumption is made which is waiting for examination. We except phase transitions in our discussion.

---

It is well-known the uncertainty relation and Schӧdinger's equation are two foundations of quantum mechanics[1], they are in dependent of each other because one cannot be derived from the other. The relation revealed a restrictive relation of quantum fluctuations between positions and momentums. A perplexed problem is which thermodynamic state the equality of the uncertainty relation corresponds to, which thermodynamic state its inequality does to. Obviously, the equation and the relation themselves cannot solve it. The same problem is also met in the quantum statistical mechanics. The probability-density operator of a mixed ensemble is denoted by[2]

$$\rho = \sum_i \rho_i \mid \psi_i \rangle \langle \psi_i \mid, \qquad \sum_i \rho_i = 1, \qquad (1)$$

the wave function $\mid \psi_i \rangle$ of Eq.(1) is given by the Schӧdinger's equation, $\rho_i$ is the probability of subsystem of the ensemble in the state $\mid \psi_i \rangle$, $i$ takes all possible values, and each subsystem is independent of another and there is not any coherence between different states. The operator $\rho$ does not relate to the uncertainty relation directly, namely, it cannot tell us what will actually happen to the relation in the equilibrium state. In the quantum statistical mechanics the quantum fluctuations cannot be neglected, which means there must be other probability-density operator $F$ concerned in the relation, which also describes the equilibrium state, $F$ and $\rho$ will become two foundations of the quantum statistical mechanics as if the equation and the relation were in the quantum mechanics.

The second law of thermodynamics has been regarded as a macroscopic law since it was found，its microscopic origin and corresponding principle in the quantum mechanics have not been obtained. Some authors tried study this subject from the dynamic point of view， but have never achieved a certain and satisfactory conclusion[3]. In fact，the process from the non-equilibrium state (time is ordered) to the equilibrium state (time is disordered) is a mutation，which cannot be solved by means of dynamic equations. Expanding $S$ in powers of random variables about the equilibrium-state entropy $S^0$， Einstein[2] obtained Gaussian distribution of the fluctuations; Prigogine derived the minimal entropy-production principle with the same method[4]. Since $S$ is non-equilibrium–state entropy both of theories merely suit to the non-equilibrium-state fluctuations. The theories pointed out the transitions of entropies are linked to the fluctuations. We think the second law of thermodynamics maybe results from the fluctuations. According to Landau's explanation Einstein's theory cannot be applied to the quantum statistical mechanics for it to neglect quantum effects[2], and the variables' deviation only corresponds to the non-equilibrium state in his theory. It is proved the uncertainty relation is topologically invariant[5], which reminds us that the invariance of the minimal uncertainty relation maybe relates to the equilibrium-state entropy. Thus, we will start off with the uncertainty relation to discuss the fluctuations, except phase transitions.

For one-dimension, mean square-deviations of the state $|\psi_i\rangle$ are

$$(\Delta x_i)^2 = \langle (x - \langle x_i \rangle)^2 \rangle , \qquad (\Delta p_{x_i})^2 = \langle (p_x - \langle p_{x_i} \rangle)^2 \rangle \qquad (2)$$

$\langle x_i \rangle$ and $\langle p_{x_i} \rangle$ are its average values of positions and momentum. The statistical fluctuations of random variables about the average values have the following common characters: On the one hand, any quantum system, no matter what its Hamiltonian operator is, obeys the uncertainty relation,

$$(\Delta x_i)^2 (\Delta p_{x_i})^2 \geq (\frac{h}{4\pi})^2 \qquad (3)$$

$(\Delta x_i)^2$ and $(\Delta p_{x_i})^2$ are given by Eq.(2), the equality of Eq.(3) is named the minimal uncertainty relation. On the other hand, the wave function describes the statistical behavior of the large number of particles, although some specific particles' behavior violates the properties of the wave function. Let a subsystem be in the state $|\psi_i\rangle$ which particles' number be $N_i$, $x$ and $p_x$ be the position and momentum of the state. Since the wave-function equation is distinct from a particle's dynamic equation to describe the particle's moving orbits and the wave function $|\psi_i\rangle$ only has statistical meaning, $x$ and $p_x$ are the variables of the function $|\psi_i\rangle$ and they are not a specific particle's position and momentum. In such a subsystem each actual particle has itself specific position and momentum caused by various reasons: collisions, transitions among energies levels, interactions of electrons with atomic nuclei, interactions between electrons, etc. Each reason results in a specific fluctuation of position or of momentum. How can the fluctuations $(x - \langle x_i \rangle)$ and $(p_x - \langle p_{x_i} \rangle)$ represent these fluctuations? Because the number of the particles is very large all of these fluctuations can take place at the same time for a particle, and one specific fluctuation cannot be distinguished from another specific fluctuation by means of Schödinger's equation and the wave function $|\psi_i\rangle$. Therefore, the fluctuations $(x - \langle x_i \rangle)$ and $(p_x - \langle p_{x_i} \rangle)$ must be the statistical configurations of these fluctuations, namely, the fluctuations $(x - \langle x_i \rangle)$ and $(p_x - \langle p_{x_i} \rangle)$ must be thought of as the sum over these fluctuations caused by various reasons. Considering that a fluctuation caused by one of these reasons is independent of another, we can say the fluctuations $(x - \langle x_i \rangle)$ and $(p_x - \langle p_{x_i} \rangle)$ obey the central limit theorem, and the fluctuations accord with Gaussian distribution[6,7]:

$$f_i(x, p_x) = \frac{1}{2\pi \Delta x_i \Delta p_{x_i}} \exp\{-\frac{1}{2}[\frac{(x - \langle x_i \rangle)^2}{(\Delta x_i)^2} + \frac{(p_x - \langle p_{x_i} \rangle)^2}{(\Delta p_{x_i})^2}]\} \qquad (4)$$

$\Delta x_i$ and $\Delta p_{x_i}$ are given by Eq.(2), and they obey the uncertainty relation. Being different from the traditional fluctuations' theory, taking quantum effects into account, the fluctuations can exist in the equilibrium state, which means that the fluctuations will not change the microscopic-states number so that the entropy of the system still is the greatest. We noticed $f_i(x, p_x)$ and Einstein's formula of fluctuations have the same form of function, an important difference between them is that the variables $x$ and $p_x$ of $f_i(x, p_x)$ only correspond to the equilibrium state, the variables $x$ and $p_x$ of Einstein's formula only do to the non-equilibrium state, which means while Einstein's formula is valid the number of microscopic states will change to turn the system's entropy into smaller. The difference between Eq.(4) and Einstein's formula is determined by the value of $\Delta x_i \Delta p_{x_i}$ which affects the probability density of the fluctuation distribution.

At first, we analyze the fluctuations qualitatively. Suppose that a system is in the equilibrium state, but the inequality of the uncertainty relation is established. It is clear that in the non-equilibrium state far away from the equilibrium state the fluctuations are very great which the inequality of the uncertainty relation satisfies, but the system do be in the non-equilibrium state. Even if in the neighborhood of the equilibrium state the entropy of system itself is a Lyapounov function to the isolated system, which makes the fluctuations decrease to the smallest and turns the system's state into the equilibrium state[7]. Thus, the case is impossible, and the equality of the uncertainty relation should be valid in the equilibrium state.

When $x = \langle x_i \rangle$, $p_x = \langle p_{x_i} \rangle$, $f_i(x, p_x)$ of Eq. (4) takes the form

$$f_i(\langle x_i \rangle, \langle p_{x_i} \rangle) = \frac{1}{2\pi \Delta x_i \Delta p_{x_i}} \qquad (5)$$

Since the fluctuations in the equilibrium state are the smallest, Eq.(5) should be the greatest, which guarantees the subsystem to be in a statistical average state, $x = \langle x_i \rangle$ and $p_x = \langle p_{x_i} \rangle$, for the longest time to meet the requirements of the ensemble theory. Obviously, only the minimal uncertainty relation can lead to this situation, which means the minimal uncertainty relation corresponds to the equilibrium state. This conclusion is in accord with the above qualitative analysis.

The particles' number of the equilibrium-state fluctuations is greater than the particles' number of the non-equilibrium-state fluctuations in the same area nearby $\langle x_i \rangle$ and $\langle p_i \rangle$, but the situation is converse in an area far away from $\langle x_i \rangle$ and $\langle p_{x_i} \rangle$. As $(\Delta x_i)^2$ and $(\Delta p_{x_i})^2$ are changeable, and what is their regular pattern? Using the minimal uncertainty relation: $(\Delta p_{x_i})^2 = [h/(4\pi \Delta x_i)]^2$, while $x \neq \langle x_i \rangle, p_x \neq \langle p_{x_i} \rangle$ and $x$, $p_x$ are very close to $\langle x_i \rangle, \langle p_{x_i} \rangle$ and are temporally considered as constant for the change of $(\Delta x_i)^2$, let the first partial derivative of $f_i(x, p_x)$ with respect to $(\Delta x_i)^2$ be zero, which indicates that $f_i(x, p_x)$ still keeps an extreme value to make the fluctuations the smallest, we then have

$$(\Delta x_i)^2 = \frac{h}{4\pi} \left| \frac{x - \langle x_i \rangle}{p_x - \langle p_{x_i} \rangle} \right| \quad , \quad (\Delta p_{x_i})^2 = \frac{h}{4\pi} \left| \frac{p_x - \langle p_{x_i} \rangle}{x - \langle x_i \rangle} \right| \tag{6}$$

Since the second partial derivative of $f_i(x, p_x)$ with respect to $(\Delta x_i)^2$ is negative while Eq.(6) is valid, it has a maximum. With the same reason for the ensemble:

$$F(x, p_x) = \frac{1}{2\pi \Delta x \Delta p_x} \exp\left\{ -\frac{1}{2} \left[ \frac{(x - \langle x \rangle)^2}{(\Delta x)^2} + \frac{(p_x - \langle p_x \rangle)^2}{(\Delta p_x)^2} \right] \right\} \tag{7}$$

$$(\Delta x)^2 (\Delta p_x)^2 \geq \left( \frac{h}{4\pi} \right)^2 \tag{8}$$

$$(\Delta x)^2 = \frac{h}{4\pi} \left| \frac{x - \langle x \rangle}{p_x - \langle p_x \rangle} \right| \quad , \quad (\Delta p_x)^2 = \frac{h}{4\pi} \left| \frac{p_x - \langle p_x \rangle}{x - \langle x \rangle} \right| \tag{9}$$

In the diagonal representations of the normalized $\rho$: $Tr(\rho) = 1$,

$$\langle x \rangle = Tr(\rho x) \quad , \qquad \langle p_x \rangle = Tr(\rho p_x) \tag{10}$$

$$(\Delta x)^2 = Tr[\rho(x - \langle x \rangle)^2] \quad , \qquad (\Delta p_x)^2 = Tr[\rho(p_x - \langle p_x \rangle)^2] \tag{11}$$

The probability density $F(x, p_x)$ conforms to the minimal uncertainty relation, describing the ensemble's fluctuations. It is important that the number of microscopic states will not change while Eq.(7) is established, which means the probability $\rho_i$ of Eq.(1) will not change for all possible states, being distinct from the traditional ensemble-fluctuations theory in which the fluctuations will change the microscopic states of the ensemble and will turn the ensemble state into the non-equilibrium state. It is interesting that the changes of $(\Delta x)^2$ and $(\Delta p_x)^2$ [or of $(\Delta x_i)^2$ and $(\Delta p_{x_i})^2$] are adjusted by the absolute value of ratio of $(x - \langle x \rangle)$ to $(p_x - \langle p_x \rangle)$ [or $(x - \langle x_i \rangle)$ to $(p_x - \langle p_{x_i} \rangle)$], obeying the minimal uncertainty relation, although they are very average values. When the inequality of the uncertainty relation is established, namely, $\Delta x \Delta p_x$ and $\Delta x_i \Delta p_{x_i}$ become greater, $F(x, p_x)$ and $f_i(x, p_x)$ become smaller for the same $\langle x \rangle$, $\langle p_x \rangle$ and $\langle x_i \rangle$, $\langle p_{x_i} \rangle$, which means the fluctuations are amplified and the ensemble's state and the subsystem's state are turned into the non-equilibrium state. Obviously, SchÔdinger's equation and $\rho$ cannot interpret these characters. So far, we proved the equality of the uncertainty relation corresponds to the equilibrium state (the entropy $S^0$), the inequality does to the non-equilibrium state (the entropy $S$), Eq.(8) and the formula of the second law of thermodynamics, $S^0 \geq S$, are one-to-one, and the uncertainty relation is a quantum-mechanics expression of the law. Using the minimal uncertainty relation and substituting Eqs.(6) and (9) in Eqs.(4) and (7) respectively, we obtain brief expressions:

$$f_i(x, p_x) = \frac{2}{h} \exp[-\frac{4\pi}{h} \mid (x - \langle x_i \rangle)(p_x - \langle p_{x_i} \rangle) \mid] \tag{12a}$$

$$F(x, p_x) = \frac{2}{h} \exp[-\frac{4\pi}{h} \mid (x - \langle x \rangle)(p_x - \langle p_x \rangle) \mid] \tag{12b}$$

Equations (12a) and (12b) are to say, the fluctuations have curves of constant distributions although $(\Delta x_i)^2$, $(\Delta p_{x_i})^2$, $(\Delta x)^2$ and $(p_x)^2$ all are changeable at the same time. When the minimal uncertainty relations in Eqs.(3) and (8) are valid, the minimal uncertainty relations of time-energy are hold[1,8]:

$$\Delta E_i \Delta t_i = \frac{h}{4\pi}, \qquad \Delta E \Delta t = \frac{h}{4\pi} \tag{13}$$

$$(\Delta E_i)^2 = \langle \psi_i \mid (E - \langle E_i \rangle) \mid \psi_i \rangle, \quad (\Delta E)^2 = Tr[\rho(E - \langle E \rangle)^2] \tag{14}$$

$$\langle E_i \rangle = \langle \psi_i \mid E \mid \psi_i \rangle, \qquad \langle E \rangle = Tr(\rho E) \tag{15}$$

Being different from $x$ (or $x_i$), $p_x$ (or $p_{x_i}$) and $E$ (or $E_i$) which are the ensemble's variables (or the subsystem's variables), $t$ and $t_i$ are belong to a reference system, which is out of the ensemble[1,8], and they act as reference variables. Since $\Delta E_i$ and $\Delta E$ are the energy fluctuations of the subsystem and the energy fluctuations of the ensemble respectively in the equilibrium state, $\Delta t$ is the interval of time while the ensemble's energy undergoes fluctuations $\Delta E$, and it is also the undergoing time of the equilibrium state. Because the subsystem is in the ensemble, $\Delta t_i$ should equal to $\Delta t$, we furthermore obtain

$$\Delta E_i = \Delta E = \frac{h}{4\pi\Delta t} = \frac{h}{4\pi\Delta t_i} \tag{16}$$

Equation (16) indicates $\Delta E_i$ and $\Delta E$ have self-similarity, which cannot be amplified; so do $\Delta x_i \Delta p_{x_i}$ and $\Delta x \Delta p_x$. While there is a continuous phase transition, the self-similarity of correlation length is resulted from fluctuations, which are amplified to the infinite $[6,7]$. Both kinds of fluctuations are two distinct limit situations.

As the minimal uncertainty variables are the fluctuations of the equilibrium state, when a system's state is measured by an apparatus, the system is disturbed and the system's state is changed to the non-equilibrium state so that the minimal uncertainty variables are unmeasured, which are considered as limits of measured values.

Finally, we suppose that because Eqs. (6) and (9) become invalid while $x = \langle x_i \rangle$, $p_x = \langle p_{x_i} \rangle$ and $x = \langle x \rangle$, $p_x = \langle p_x \rangle$, and $\Delta x_i$, $\Delta p_{x_i}$ and $\Delta x$, $\Delta p_x$ will not equal to zero forever, they must keep the smallest simultaneously so that the sum $(\Delta x_i + \Delta p_{x_i})$ and the sum $(\Delta x + \Delta p_x)$ must be the smallest restricted by the minimal uncertainty relation, which leads to results: $\Delta x_i = \Delta p_{x_i} = \frac{1}{2}(h/\pi)^{1/2}$ while $x = \langle x_i \rangle$ and $p_x = \langle p_{x_i} \rangle$ ; $\Delta x = \Delta p_x = \frac{1}{2}(h/\pi)^{1/2}$ while $x = \langle x \rangle$ and $p_x = \langle p_x \rangle$. The assumption is waiting for examination.

---